# Enhancing Petrophysical Studies with Machine Learning: A Field Case Study on Permeability Prediction in Heterogeneous Reservoirs

*Fethi Ali Cheddad*

*University of science and technology Houari Boumediene*

*fcheeddad@usthb.dz*

**Abstract:** This field case study aims to address the challenge of accurately predicting petrophysical properties in heterogeneous reservoir formations, which can significantly impact reservoir performance predictions. The study employed three machine learning algorithms, namely Artificial Neural Network (ANN), Random Forest Classifier (RFC), and Support Vector Machine (SVM), to predict permeability log from conventional logs and match it with core data. The primary objective of this study was to compare the effectiveness of the three machine learning algorithms in predicting permeability and determine the optimal prediction method. The study utilized the Flow Zone Indicator (FZI) rock typing technique to understand the factors influencing reservoir quality. The findings will be used to improve reservoir simulation and locate future wells more accurately. The study concluded that the FZI approach and machine learning algorithms are effective in predicting permeability log and improving reservoir performance predictions.

## 1. Introduction

The correlation between porosity and permeability in heterogeneous reservoir formations is generally poor (Ifrene et al., 2022; Irofti et al., 2022), posing a challenge for accurately describing reservoir properties in petrophysical studies (Abes et al., 2021; Ifrene et al., 2023; Khetib et al., 2023; Pothana et al., 2023). This can significantly impact reservoir performance predictions, highlighting the importance of accurately predicting petrophysical properties when building a representative reservoir simulation model.

To address this challenge, a field case study was conducted in the BERKINE gas-producing Algerian Shally-Sand formation using the Flow Zone Indicator (FZI) rock typing technique to understand the factors influencing reservoir quality. The study employed three machine learning algorithms, namely Artificial Neural Network (ANN), Random Forest Classifier (RFC), and Support Vector Machine (SVM), to predict permeability Log from conventional logs and match it with core data.

The primary objective of this study was to compare the effectiveness of the three machine learning algorithms in predicting permeability and determine the optimal prediction method. The findings will be used to improve reservoir simulation and locate future wells more accurately. By developing a better understanding of the relationship between porosity and permeability in heterogeneous formations, petrophysicist can make more informed decisions and ultimately enhance reservoir performance.

Despite advances in petroleum engineering technology, there are no logging tools available that can provide continuous permeability log measurement. Therefore, this study focused on estimating the permeability log from existing plugs in cored wells and predicting this rock property in uncored wells or intervals using a supervised classification method of rock types obtained from a statistical approach. A comparison between ANN, RFC, and SVM machine learning algorithms was made to define the most accurate model to use in this field.

## 2. Methods

To address the common challenge of predicting permeability in uncored but logged wells, the hydraulic flow unit (HFU) approach will be utilized for the classification of rock types and prediction of flow properties. The HFU is defined as a representative elementary volume of total reservoir rock with internally consistent geological and petrophysical properties that control fluid flow and are predictably different from properties of other rocks.

The fundamental petrophysical units in a reservoir, or rock types, can be determined by flow zone indicators for routine core plug analysis. These petrophysical properties, such as porosity and permeability, should have small variations for a given rock type, indicating that knowledge of any one property can enhance the prediction of the other. The FZI can be calculated from core data using the normalized porosity index (NPI) and reservoir quality index (RQI) through equation 2, which was introduced by Amaefule et al. in 1993.

$$K = \frac{\emptyset_e^3}{(1-\emptyset_e)^2} \frac{1}{F_s \tau^2 S_{gv}^2}$$

$$0.0314 \sqrt{\frac{K}{\emptyset_e}} = \left(\frac{\emptyset_e}{1-\emptyset_e}\right)\left(\frac{1}{\sqrt{F_s}\tau S_{gv}}\right)$$

K : Permeability in md

$\emptyset_e$ : Effective Porosity in fraction

Fs : Shape fraction   $\tau$ : Tortuosity

Sgv : Surface area per unit pore volume in µm-1

FZI : Flow Zone Indicator in µm

The Flow Zone Indicator (FZI) is a valuable and distinct value that quantifies the flow behavior of a reservoir and offers a relationship between small-scale petrophysical properties, like core plugs, and large-scale properties, such as those observed at the well bore level. FZI represents flow zones based on surface area and tortuosity, making it a useful term. It can be mathematically represented as shown in the study conducted by Al-Dhafeeri et al. in 2007.

Once the FZI has been calculated for each well, it is necessary to integrate the porosity and permeability in ascending order to determine the normal probability of each FZI sample. The objective of this step is to identify the number of rock types. A change in tangent in the curve (NormalProbability-logFZI) signifies a change of rock type, as depicted in Figure 1. Figure 2 and Figure 3 show the distinction of rock types based on different properties.

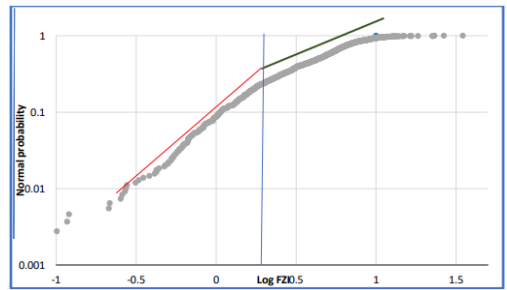

Figure 1: Normal Probability-FZI log Curve

| FZI Mean | Rock Type |
|---|---|
| 1,314080176 | HFU1 |
| 6,149094233 | HFU2 |

Table 1: Number of Rock Types Determination Via FZI Mean

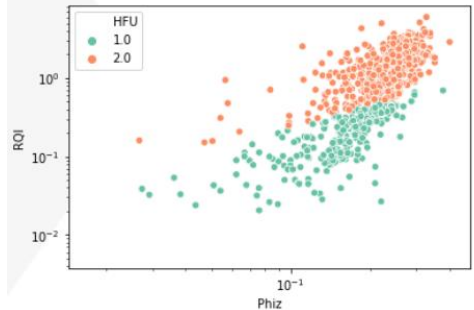

Figure 2: Rock Types distinction from RQI-Phiz Scatter plot

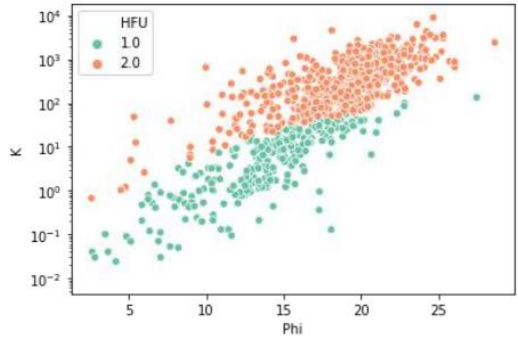

Figure 3: Rock Types distinction from Permeability-Porosity Scatter plot

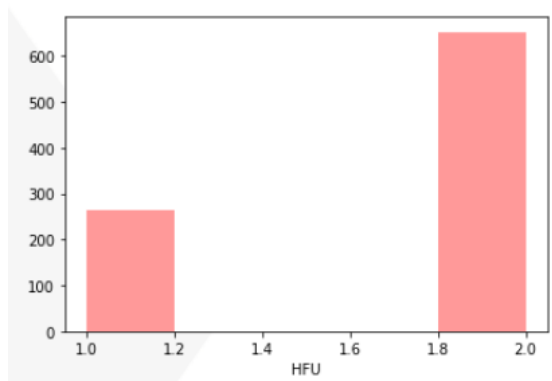

Figure 4: Rock Types distribution in the available samples

Figure 4 highlights an imbalanced distribution of rock types in the dataset. This can lead to bias that affects the quality of the model. It is worth noting that the rock types classes were obtained after the data cleaning step in the data preprocessing stage.

## 2.1. Data Preprocessing

Data preprocessing is a crucial stage in the data analysis pipeline as it involves transforming raw data into a format that is suitable for analysis. The process consists of several steps, including data cleaning, integration, visualization, transformation, and reduction. These steps aim to remove any inconsistencies, errors, or missing values in the data, combine data from multiple sources, and reduce the size of the dataset to improve computational efficiency.

Various software tools are used to execute these tasks, including Microsoft Excel, TechLog, and Python. Microsoft Excel is a popular tool for data cleaning and transformation, while TechLog is a specialized software used in the oil and gas industry for log data analysis. Python is a general-purpose programming language widely used in data science and machine learning for its powerful data manipulation and analysis libraries. By performing these preprocessing steps, data scientists can ensure that the data they analyze is accurate, reliable, and suitable for the intended purpose.

### 2.1.1. Data Cleaning

One crucial step in any data mining project is to clean the dataset by addressing missing values, noisy data, and inconsistent data. In other words, if you input garbage, you will get garbage output. Due to the sensitivity of the core permeability-porosity dataset, the cleaning process is done manually using excel. This involves removing cracked plugs with high permeability that do not accurately represent the rock's true nature, as well as dealing with missing porosity and permeability values in different wells.

To ensure accuracy, porosity and permeability measurements are sometimes taken two or three times, but only the first measurement is used to avoid redundancy resulting from plug manipulation. The dataset's inputs for permeability prediction include depth, density, gamma ray, sonic, neutron, porosity, and permeability, with the last two obtained from plugs.

In this particular study, the dataset is reduced from 1149 raw examples to 918 for predicting permeability. One well's dataset is excluded due to the lack of neutron and density inputs, while another well's dataset is included for determining rock types, where only FZI is needed. This inclusion reduces the total number of examples to 1087 raw, not 918.

After cleaning the datasets, the next step is to integrate all the datasets from different wells into a single dataset. This integration is necessary to prepare the dataset for the rock typing classification, which is used as the label for the supervised classification method.

Data integration is a critical step in many data analysis projects, as it can lead to problems with redundancy. However, in this study, redundancy is not an issue as the datasets are carefully curated and the data cleaning process has removed any redundant or irrelevant information. Therefore, the integrated dataset is a high-quality representation of the relevant information from all the wells and is ready for further analysis.

### 2.1.2. Data Visualization

Data visualization is a method of representing information and data through visual elements such as charts, graphs, and maps. These tools offer an intuitive and accessible way to observe and comprehend patterns, outliers, and trends within the data. By presenting information visually, data visualization enables individuals to make more decisions that are informed and identify insights that may have otherwise gone unnoticed.

Table 2 shows a descriptive statistics including those that summarize the central tendency dispersion and the shape of the whole dataset.

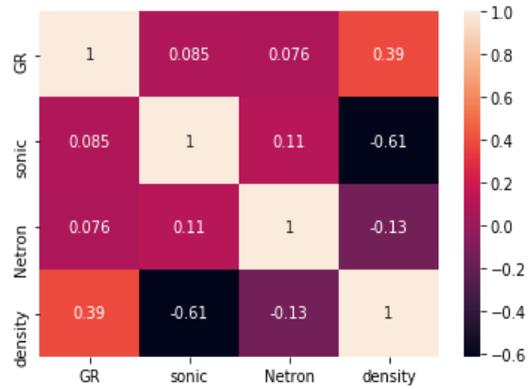

Figure 5: Heat Map shows the correlation between all the features

The Heat Map in figure 6 shows a relatively high correlation coefficient (inversely proportional) between sonic & density inputs. On the other hand, all the other features have lower correlation coefficients.

|  | Depth | density | GR | sonic | Neutron | K | Phi | Phiz | RQI | FZI | HFU |
|---|---|---|---|---|---|---|---|---|---|---|---|
| count | 917.000000 | 917.000000 | 917.000000 | 917.000000 | 917.000000 | 917.000000 | 917.000000 | 917.000000 | 917.000000 | 917.000000 | 917.000000 |
| mean | 3052.927035 | 2.400153 | 59.856023 | 75.155857 | 7.263788 | 402.812517 | 17.497928 | 0.215144 | 1.085542 | 4.633375 | 1.711014 |
| std | 112.053723 | 0.102178 | 24.065365 | 4.420714 | 8.488768 | 684.191498 | 4.215062 | 0.059964 | 0.866158 | 3.254610 | 0.453539 |
| min | 2929.280000 | 2.196958 | 16.080000 | 55.433100 | 0.082200 | 0.024000 | 2.600000 | 0.026694 | 0.020469 | 0.120405 | 1.000000 |
| 25 % | 2965.704000 | 2.327326 | 43.446700 | 72.326500 | 0.152000 | 22.300000 | 14.920000 | 0.175364 | 0.374254 | 1.981864 | 1.000000 |
| 50 % | 2999.080000 | 2.380000 | 57.633800 | 75.374600 | 0.197100 | 153.120000 | 18.330000 | 0.224440 | 0.931742 | 4.217611 | 2.000000 |
| 75 % | 3173.273000 | 2.463400 | 73.505000 | 78.058600 | 16.270000 | 504.599900 | 20.530000 | 0.258336 | 1.574537 | 6.420537 | 2.000000 |
| max | 3285.896000 | 2.895300 | 178.492000 | 93.025300 | 26.468600 | 9010.163000 | 28.610000 | 0.400756 | 6.003265 | 23.209655 | 2.000000 |

Table 2: Dataset Description

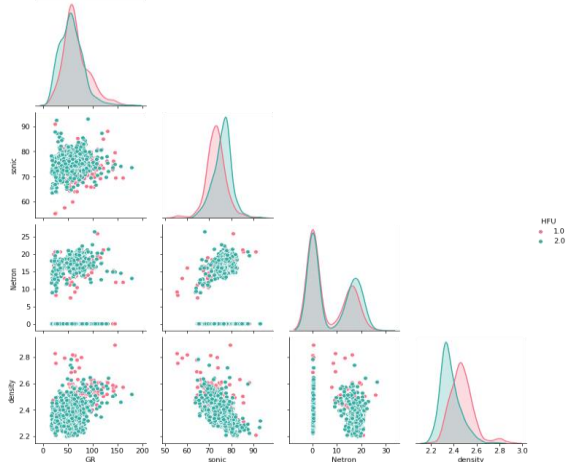

Figure 6: Matrix Plot shows the relationship between all the features

### 2.1.3. Data Transformation

Data transformation is the process of changing the format, structure, or values of data to be appropriate for mining, in order to get better results It can be realized in several methods (Smoothing, Aggregation, Attribute Construction and Normalization). Normalization is the Data transformation method used for scaling in this project

- *Normalization*

The attribute data are scaled to fail within a small specified range for the purpose of making the dataset adapted to the mining task, a scaling is used in order to get the same range between features.

It performs a linear transformation on the original data depending on the maximum and minimum values of the attribute and the new range values.

$$z = \frac{x - minA}{maxA - minA} * (nmaxA - nminA) + nminA$$

Where, minA and maxA are the minimum and maximum values of attribute A respectively. nminA and nmaxA are the new minimum (0) and new maximum (1) values of attribute A respectively.

### 2.1.4. Data Reduction

Data reduction is a technique that can provide a more concise representation of the original data while preserving its quality. This is achieved by selecting or transforming relevant features to create a smaller dataset that still contains valuable information.

The selection of the appropriate method for data representation, selection, reduction, or transformation of features is crucial in determining the quality of the data mining outcome. One of the most critical operations in a data reduction process is feature selection, which involves removing irrelevant attributes from the dataset. By reducing the number of features, feature selection can simplify the data mining process, improve the accuracy of predictive models, and reduce the risk of overfitting.

Feature selection is a technique used to identify the relevant features in a dataset that are necessary for a specific data mining application. The objective is to achieve maximum performance with minimal measurement and processing effort by reducing the dataset size through the removal of irrelevant or redundant attributes.

There are two types of feature selection techniques: unsupervised and supervised. Unsupervised techniques reduce the number of fields without considering the class label or class target, while supervised techniques select features based on the class label.

In this case study, the depth feature is considered irrelevant for the prediction of permeability and has been manually removed from the dataset. Other features may also be removed based on their relevance to the dataset. The feature selection process was conducted in an unsupervised manner, where the label was not considered, using the Principal Component Analysis (PCA) technique. PCA identifies the most significant features in the dataset by reducing the dimensionality of the data while retaining the maximum amount of information possible.

- **Principal Component Analysis (PCA)**

Principal Component Analysis (PCA) is a data reduction technique that reduces the dimensionality of datasets while preserving the essential information. It achieves this by generating new variables that are uncorrelated with one another and successively maximize variance.

PCA is useful for increasing the interpretability of datasets by reducing the complexity of the data without significant information loss. This is especially important for large datasets where visualization and analysis can become challenging due to the high number of variables. By creating new variables that explain the maximum amount of variance in the dataset, PCA can help researchers identify the most significant features that contribute to the underlying patterns in the data.

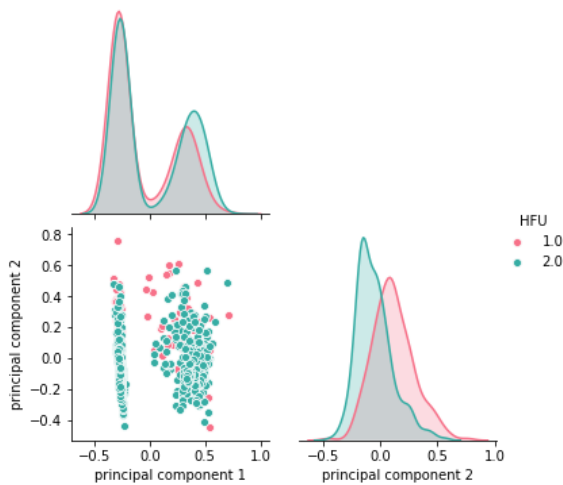

Figure 7: Matrix Plot shows the correlation between the two principal components

For analyzing the most efficient case when it comes to the chosen number of principal components. A three and two PCA is tried, and it gave the same results in terms of model efficiency. So, to optimize the model, only two principal components are taken into consideration.

- **Under Sampling**

Initially, the data mining process was performed without considering the issue of unbalanced class distribution, resulting in a poor model in terms of sensitivity, precision, and F1 score due to a bias towards the class with more samples. To address this problem, either an oversampling technique, which involves randomly increasing the number of samples in the class with fewer samples to match that of the majority class, or an undersampling technique, which involves reducing the number of samples in the majority class to match the minority class, can be applied. However, as the permeability values being predicted in this project are highly sensitive, the use of an oversampling technique may result in erroneous results. Thus, the undersampling method was employed, which significantly improved the quality of the results.

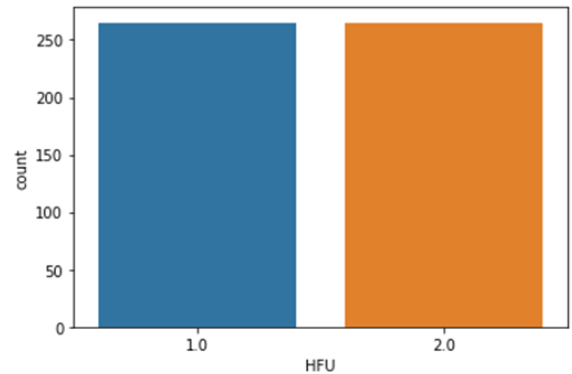

Figure 8: The distribution of classes after Under Sampling

Following the application of the under sampling technique, the original number of samples in the majority class, class two, was reduced from 652 to 256 samples, to equalize the sample size with the minority class.

### 2.2. Data Mining

Machine learning algorithms are used for data mining tasks, which can be categorized into three types:

- Descriptive Analytics (Unsupervised Learning): Identify underlying patterns or structure in the historical data.
- -Predictive Analytics (Supervised Learning): Learn a model to predict Y from X.
- Prescriptive Analytics: Mix of descriptions and predictions to prescribe or recommend the next set of actions to be done.

The objective of this study is to predict the rock type of each example to predict the permeability log. Therefore, it involves a classic data mining task of supervised learning using classification algorithms such as Artificial Neural Networks (ANN), Random Forest, and Support Vector Machine (SVM). The aim is to determine the best algorithm for the prediction by defining the goodness of each model.

The machine learning classifier consists of two stages:

- Training set: It takes training data, which is a set of data points with their corresponding correct labels, and tries to learn a pattern for how the points map to the label. This stage represents 80% of the total dataset.
- Test set: Once the classifier is trained, it acts as a function that takes in additional data points and outputs predicted classifications for them. The prediction is a specific label. This stage represents 20% of the total dataset without labels.

The three classification algorithms used in this study are ANN, Random Forest, and SVM.

### 2.2.1. Artificial Neural Network (ANN)

One way to implement machine learning is through Artificial Neural Networks (ANN). ANN consists of input and output units connected with weights that are adjusted during the learning phase to predict the correct class label of input tuples. The process starts by specifying the number of units in the input layer, followed by determining the number of hidden layers and units in each hidden layer, and finally the output layer with the number of classes.

For this case study, an ANN with 4 hidden layers was used. The number of units in each hidden layer are 5, 4, 3, and 2 respectively. The optimization process was carried out over 100,000 iterations using LBFGS as the solver.

- **Activation Function (Transfert Function)**

Several activation functions were used, ReLu gave the best results in terms of accuracy. It is a non-linear activation function that is used in multi-layer neural networks or deep neural networks. This function can be represented as:

$$f(x) = \begin{cases} 0, if\ x < 0 \\ x, if\ x \geq 0 \end{cases}$$

### 2.2.2. Support Vector Machine (SVM)

In this study, SVM is used as a supervised learning algorithm to analyze the data used for classification. It constructs a hyperplane with a good separation achieved by the hyperplane having the largest distance to the nearest training-data point of any class, known as the functional margin. A larger margin results in a lower generalization error of the classifier. SVM is chosen due to the number of dataset examples available, making it one of the most suitable algorithms. To handle the unbalanced distribution of the samples in each class, the Kernel sigmoid function is used as an argument in addition to the balanced class weight in this model.

### 2.2.3. Random Forest Classifier (RFC)

Random forest is a popular supervised learning algorithm that constructs multiple decision trees using randomly selected data samples. Each tree makes a prediction, and the final solution is chosen based on the collective predictions of all the trees. This method is known as voting. Random forest is chosen for modeling in this study because it is known to be highly accurate and robust, particularly when many decision trees are used. In this study, 300 estimators, or decision trees, are used to create the random forest model.

## 2.3. Classification Model Evaluation

The evaluation of a classification model involves measuring various aspects such as accuracy, speed, robustness, scalability, and interpretability. While accuracy is the primary metric used for evaluation, there are other important metrics to consider as well. To assess the performance of each model, the confusion matrix is used. This matrix allows for the calculation of various classification parameters in Python to evaluate the overall power of the model.

### 2.3.1. Classification Model Evaluation for ANN

|  |  | Predicted Class |  |  |
|---|---|---|---|---|
|  |  | HFU1 | HFU2 |  |
| Tested Class | HFU1 | 36 (TP) | 15 (FN) | 51 (P) |
|  | HFU2 | 12 (FP) | 43 (TN) | 55 (N) |
|  |  | 48 (P') | 58 (N') | 106 (all) |

Table 3: Confusion Matrix of ANN Modeling

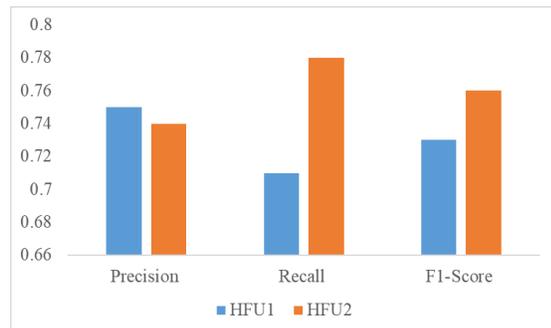

Figure 9: Precision, Recall, F1-Score of ANN

| Accuracy |  |  | 0.75 | 106 |
|---|---|---|---|---|
| Macro avg | 0.75 | 0.74 | 0.74 | 106 |
| Weighted avg | 0.75 | 0.75 | 0.74 | 106 |

Table 4: Shows the Accuracy results of ANN

### 2.3.2. Classification Model Evaluation for SVM

|  |  | Predicted Class |  |  |
|---|---|---|---|---|
|  |  | HFU1 | HFU2 |  |
| Tested Class | HFU1 | 36 (TP) | 15 (FN) | 51 (P) |
|  | HFU2 | 13 (FP) | 42 (TN) | 55 (N) |
|  |  | 49 (P') | 57 (N') | 106 (all) |

Table 5: Confusion Matrix of SVM

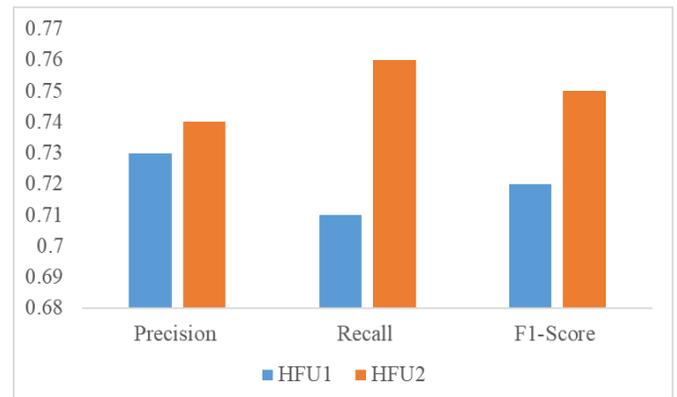

Figure 10: Precision, Recall, F1-Score of SVM

| Accuracy |  |  | 0.74 | 106 |
|---|---|---|---|---|
| Macro avg | 0.74 | 0.73 | 0.73 | 106 |
| Weighted avg | 0.74 | 0.74 | 0.74 | 106 |

Table 6: Shows the Accuracy results of SVM

### 2.3.3. Classification Model Evaluation for RFC

|  |  | Predicted Class | |  |
|---|---|---|---|---|
|  |  | HFU 1 | HFU 2 |  |
| Tested Class | HFU 1 | 41 (TP) | 14 (FN) | 55 (P) |
|  | HFU 2 | 17 (FP) | 34 (TN) | 51 (N) |
|  |  | 58 (P') | 48 (N') | 106 (all) |

Table 7: Confusion Matrix of RFC

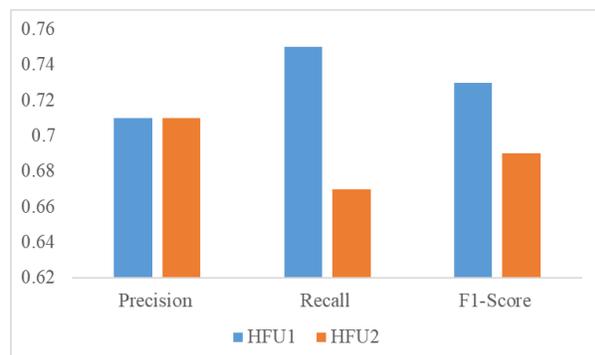

Figure 11: Precision, Recall, F1-Score of RFC

| Accuracy |  |  | 0.71 | 106 |
|---|---|---|---|---|
| Macro avg | 0.71 | 0.71 | 0.71 | 106 |
| Weighted avg | 0.71 | 0.71 | 0.71 | 106 |

Table 8: Shows the Accuracy results of RFC

### 2.3.4. Models Comparison

In comparing the models, it can be observed that the ANN yields the highest accuracy with 75%, followed by SVM with 74%, while Random Forest has the lowest accuracy with only 71%. However, accuracy alone is not the sole determinant of the best model. Other parameters such as precision, recall, and F1 score also play a crucial role in assessing the model's goodness. In this study, ANN produced the best results for precision, recall, and F1 score with 0.75, 0.74, and 0.74 respectively. SVM is the second-best algorithm for classification in this study. In contrast, Random Forest produced the weakest results.

After constructing and assessing the models, a new well that was not previously used to train the ANN, SVM, and RFC models was chosen to test the generalizability of the results. This process is commonly referred to as knowledge extraction.

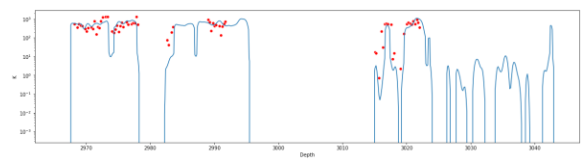

Figure 12: K_Log predicted from ANN & Permeability Core from well VS Depth

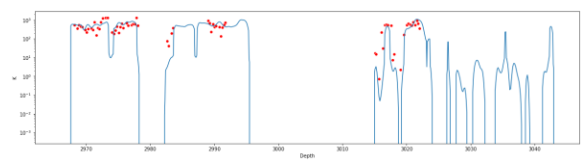

Figure 13: K_Log predicted from SVM & Permeability Core from well VS Depth

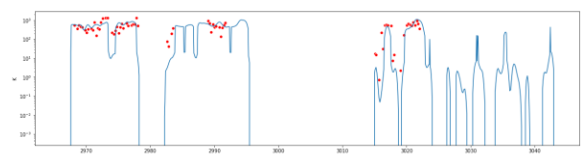

Figure 14: K_Log predicted from RFC & Permeability Core from well VS Depth

The Pearson & Spearman coefficient of correlation between the permeability predicted & the permeability in cores on the well taken for the models generalization is calculated, and it is shown in the following tables.

|  | K_core |
|---|---|
| K_ANN | 0.54 |
| K_SVM | 0.55 |
| K_RFC | 0.55 |

Table 9 : Pearson Correlation between K_Predicted & K_Core

|        | K_core |
|--------|--------|
| K_ANN  | 0.56   |
| K_SVM  | 0.56   |
| K_RFC  | 0.57   |

Table 10 : Spearman Correlation between K_Predicted & K_Core

## 3. Conclusion

This study presents a data mining approach for predicting a continuous permeability log in petroleum engineering. The dataset was prepared by performing data cleaning, integration, transformation using Min-Max normalization, and data reduction using PCA. Three machine learning algorithms, namely SVM, ANN, and RFC Classifier, were trained on the dataset. The results indicate that the performance of the models is comparable, with a slight advantage observed for the ANN model in terms of model evaluation. Additionally, rock typing using the FZI/RQI approach was employed to gain insights into factors controlling reservoir quality and fluid flow characteristics to build an accurate permeability model. The dataset available for this case study was limited to two rock types. Thus, we recommend future work to expand the dataset with more rock types to improve the accuracy of the permeability prediction. Overall, the study highlights the importance of data mining in petroleum engineering and provides valuable insights for predicting permeability logs.